\documentclass[doublespacing]{elsart_modified}

\usepackage{natbib}

\usepackage{graphicx}

\newcommand{\tempelfull}{\mbox{9P/Tempel\,1}}
\newcommand{\gs}{\mbox{$1_{10}\rightarrow 1_{01}$}}

\newcommand{\pers}{\mbox{s$^{-1}$}}
\newcommand{\kms}{\mbox{km\,s$^{-1}$}}
\newcommand{\Kkms}{\mbox{K\,km\,s$^{-1}$}}

\newcommand{\qwat}{Q_{\rm H_2O}}

\newcommand{\vexp}{v_{\rm exp}}
\newcommand{\rh}{r_{\rm h}}

\newcommand{\tkin}{T_{\rm kin}}

\newcommand{\trdCO}{\mbox{$^{13}$CO}}

\newcommand{\Jfivefour}{\mbox{J=5$\rightarrow$4}}

\newcommand{\Tmb}{T_{\rm mb}} 
\newcommand{\Imb}{I_{\rm mb}}

\newcommand{\new}{}
\newcommand{\aj}{Astron. J.} 
\newcommand{\apj}{Astrophys. J.} 
\newcommand{\apjl}{Astrophys. J. Letter} 
\newcommand{\aap}{Astron. Astrphys.}
\newcommand{\mnras}{Month. Not. Royal Astron. Soc.} 
\def\la{\mathrel{\mathchoice {\vcenter{\offinterlineskip\halign{\hfil
$\displaystyle##$\hfil\cr<\cr\sim\cr}}}
{\vcenter{\offinterlineskip\halign{\hfil$\textstyle##$\hfil\cr
<\cr\sim\cr}}}
{\vcenter{\offinterlineskip\halign{\hfil$\scriptstyle##$\hfil\cr
<\cr\sim\cr}}}
{\vcenter{\offinterlineskip\halign{\hfil$\scriptscriptstyle##$\hfil\cr
<\cr\sim\cr}}}}}
\def\ga{\mathrel{\mathchoice {\vcenter{\offinterlineskip\halign{\hfil
$\displaystyle##$\hfil\cr>\cr\sim\cr}}}
{\vcenter{\offinterlineskip\halign{\hfil$\textstyle##$\hfil\cr
>\cr\sim\cr}}}
{\vcenter{\offinterlineskip\halign{\hfil$\scriptstyle##$\hfil\cr
>\cr\sim\cr}}}
{\vcenter{\offinterlineskip\halign{\hfil$\scriptscriptstyle##$\hfil\cr
>\cr\sim\cr}}}}}

\begin{document}

\begin{frontmatter}



\title{Submillimeter Wave Astronomy Satellite observations of 
comet 9P/Tempel 1 \\ and Deep Impact}


\author[cfa,bonn]{Frank Bensch},
\author[cfa]{Gary J. Melnick},
\author[jhu]{David A. Neufeld}, 
\author[dc-cornell]{Martin Harwit},
\author[umass]{Ronald L. Snell}, 
\author[cfa]{Brian M. Patten}, 
\author[cfa]{Volker Tolls}

\address[cfa]{Harvard-Smithsonian Center for Astrophysics, 
60 Garden Street, Cambridge, MA 02138, U.S.A.}
\address[bonn]{Argelander-Institut f\"ur Astronomie, Universit\"at
Bonn, 53121 Bonn, Germany\thanksref{aifa} }
\address[jhu]{Department of Physics and Astronomy, Johns Hopkins 
University, Baltimore, MD 21218, U.S.A. }
\address[dc-cornell]{511 H street, SW, Washington, DC 20024, U.S.A.; also Cornell University}
\address[umass]{Department of Astronomy, University of Massachusetts, 
Amherst, MA 01003, U.S.A.}

\thanks[aifa]{founded in 2006 by merging of the Sternwarte, 
Radioastronomisches Institut and Institut f\"ur Astrophysik und 
Extraterrestrische Forschung der Universit\"at Bonn}



%
%
%
%
%


\end{frontmatter}



\begin{flushleft}
\vspace{1cm}
Number of pages: \pageref{lastpage} \\
Number of tables: \ref{lasttable}\\
Number of figures: \ref{lastfig}\\
\end{flushleft}


\newcommand{\sep}{; }


\pagebreak

\noindent
\textbf{Proposed Running Head:}\\
  SWAS observations of 9P/Tempel 1

\vspace{3cm}
\noindent
\textbf{Please send Editorial Correspondence to:} \\
Frank Bensch \\
Argelander Institut f\"ur Astronomie\\
Abteilung Radioastronomie\\
Auf dem H\"ugel 71\\
53121 Bonn, Germany.\\
\\
Email: fbensch@astro.uni-bonn.de\\
Phone: +49 (228) 73-1774 \\
Fax: +49 (228) 72-1775

\vfill

\pagebreak


\noindent
\textbf{ABSTRACT}

On 4 July 2005 at 5:52 UT the Deep Impact mission successfully completed
its goal to hit the nucleus of \tempelfull\ with an impactor, forming a crater 
on the nucleus and ejecting material into the coma of the comet. 
NASA's Submillimeter Wave Astronomy Satellite (SWAS) observed 
the $1_{10}-1_{01}$ ortho-water ground-state rotational transition in 
comet \tempelfull\ before, during, and after the impact.  
No excess emission from the impact was detected by SWAS and
we derive an upper limit of $1.8\times 10^7$\,kg on the water ice evaporated
by the impact. However, the water production rate of the comet showed 
large natural variations of more than a factor of three during the 
weeks before and after the impact. Episodes of increased
activity with $\qwat\sim 10^{28}$~molecule~s$^{-1}$ alternated with periods
with low outgassing ($\qwat\la 5\times 10^{27}$~molecule~s$^{-1}$). 
We estimate that \tempelfull\ vaporized a total of $N\sim 4.5\times 10^{34}$ 
water molecules ($\sim 1.3\times 10^9$\,kg) during 
{\new June-September 2005}.
Our observations indicate that only a small fraction of the nucleus 
of Tempel~1 appears to be covered with active areas. 
Water vapor is expected to emanate predominantly from 
{\new topographic features} periodically facing the Sun as
the comet rotates.  We calculate that appreciable asymmetries of these
features could lead to a spin-down or spin-up of the nucleus at observable
rates.

\vspace{\fill}

\pagebreak

\section{Introduction}
\label{intro}
The primary science goal of the Deep Impact mission is
the study of the physical and chemical properties of the material
at and below the surface of a comet.  This was accomplished by directing 
a projectile at the nucleus of comet \tempelfull. 
The 370\,kg impactor separated from its fly-by mother spacecraft one 
day before encounter with the comet and hit the sunlit side of the 
nucleus with a relative
velocity of 10.3\,\kms,  releasing a kinetic energy of 19\,GJ
at the impact site. The impact and the
ejecta were observed by the fly-by spacecraft of the Deep Impact mission
\citep{AHearn2005} as well as by a large number of Earth-based 
telescopes and satellite-based observatories
\cite[for an overview, see][]{Meech2005a}.

\tempelfull, the target comet of the Deep Impact mission, is a
low-activity Jupiter-family comet with an orbital period of
5.5\,years and a slow rotation period of $41.85\pm0.10$\,hrs measured
before the 2005 apparition \citep[see][ for a summary]{Meech2005b,Belton2005}. 
 Previous estimates of water vapor production through OH measurements 
give a typical evaporation rate of $\qwat\sim10^{28}$~molecule~s$^{-1}$
near perihelion\footnote{following the convention commonly 
used in the literature we write $\qwat$ in units of \pers\ in the 
remainder of the paper}
\citep[][]{Lisse2005}. The abundance of minor gas species 
in the coma, such as C$_2$ and CN, is at the low end of the range 
derived for 'typical' comets \citep[following the taxonomy 
by][]{AHearn1995}. The dust probed by observations
at optical wavelengths indicates that \tempelfull\ is neither
particularly dust-poor nor dust-rich.

The choice of \tempelfull\ as the target comet for the Deep Impact
mission gave rise to a large number of observing campaigns, 
which have increased our knowledge
of the physical properties of this comet, the chemical composition of
the dust and volatiles in its coma, and its periodic variations in 
activity. The observing campaign for \tempelfull\ 
culminated during the weeks leading up to the Deep Impact event 
and the days following impact.  First results from the
impact experiment have been reported by \cite{AHearn2005}, and
an overview on the results from the Earth-based observing campaign 
is given by \cite{Meech2005a}. First results constraining the
amount and composition of the dust in the ejecta are presented by 
\cite{Sugita2005} and \cite{Harker2005}, while the volatiles released 
by the impact have been studied by \cite{Kueppers2005}, \cite{Mumma2005}, 
and \cite{Keller2005}. Most of the observers
report that the activity triggered by the impact was relatively 
short lived, and that the comet (coma) returned to its pre-impact
state $\la 5$ days after the impact. 
Observers conducting monitoring observations for \tempelfull\ over
a longer period reported frequent natural outbursts of the comet
\citep[][and references therein]{Meech2005a}.
{\new With regard to the total brightness of the comet,}
these were often much larger than the outburst triggered by the impact,
but studied
in less detail than the impact itself since most of the observations
were conducted around the impact date, July 4, 2005.

Water is the most abundant species among the volatiles in cometary comae. 
Spectra of water vapor yield information on the amount of water evaporated 
from the nucleus, and provide reference levels for deriving the relative 
abundances of minor volatiles while also monitoring overall cometary activity.
Water line observations proved useful also for estimating the total amount 
of water vapor released in response to the Deep Impact event.

Most of the water is rotationally cold throughout the cometary atmosphere
\citep{Crovisier1984,Weaver1984,Bockelee-Morvan1987,Xie1992,Bensch2004b}. 
The ground-state transition 
of ortho water is therefore particularly suitable for tracing 
the water production rate.
Present-day heterodyne receivers provide the high velocity resolution
($<1$\,\kms) required to resolve cometary emission lines. Observations
of low-lying water rotational transitions, however, have to be carried out from
space-based platforms because the Earth's atmosphere is 
opaque at the line frequencies of these transitions in the submillimeter and
far-infrared regime.  

Three satellites with instruments capable of observing the 
\gs\ ground-state transition of ortho water were available during
the 2005 apparition of \tempelfull\ and have participated in the
Deep Impact observing campaign: the Microwave Instrument for the
Rosetta orbiter, MIRO, 
the Swedish-led satellite mission
Odin {\new \citep{Nordh2003},} and the Submillimeter Wave Astronomy Satellite,
SWAS {\new \citep{Melnick2000}.} 
In this paper we present the results from the SWAS observing
campaign for comet \tempelfull\ and Deep Impact. Constrained only 
by the comet's visibility from Earth orbit and a Moon-avoidance angle, 
SWAS conducted daily monitoring observations of the comet covering
approximately a 3-month period, from 5 June to 1 September 2005. 
The SWAS observations are
presented in Sect.\ \ref{swas-obs}. These observations are used in
Sect.\ \ref{water-prod} to study the variations in water vapor production
and to constrain the amounts of water ice vaporized as a result of the Deep 
Impact event.
A discussion of the results is given in Sect.\ \ref{discuss}.

\section{SWAS observations}
\label{swas-obs}
NASA's Submillimeter Wave Astronomy Satellite is a
complete space-borne radio observatory \citep{Melnick2000}. SWAS
observes the 556.936\,GHz \gs\ ground-state transition of
ortho water simultaneously with the transitions of three
other species of astrophysical interest (\trdCO\ \Jfivefour\ at
550.926\,GHz, O$_2$ at 487.249\,GHz, and  [CI] at 492.161\,GHz).
A total co-average of the data revealed no line in any of the latter species,
however. The velocity resolution of the SWAS acousto-optical spectrometer
is 0.8\,\kms. The angular size
of the elliptical SWAS beam at 556.9\,GHz is $3.3' \times 4.3'$
(full width at half maximum). This corresponds to a linear size of 
127\,000\,km $\times$174\,000\,km at $\Delta=0.89$\,AU, the comet-Earth
distance of \tempelfull\ on 4 July 2005. The SWAS beam thus
encompasses most of the gaseous water region 
in the coma of \tempelfull.

SWAS completed its highly successful mission in July 2004. On
1 June 2005 SWAS was brought out of its orbital hibernation for
the purpose of supporting the Deep Impact mission. Check-out
of the observatory conducted between 1 June and 5 June showed
no measurable degradation in the observatory's performance since 
July 2004. SWAS monitoring observations of \tempelfull\ began
on UT 5.29 June 2005 and were continued until 1.84 September 2005.
(All time data in the paper are given in UT.)
The comet-Earth distance increased from $\Delta=0.77$\,AU to
$\Delta=1.34$\,AU over this period, while the heliocentric
distance varied only slightly between $\rh=1.51$\,AU and $1.61$\,AU.
The comet reached its perihelion at $\rh=1.506$\,AU on 5.32 July 2005.

SWAS is in a 650\,km circular orbit around the Earth, and
observations were made whenever the comet was 45$^{\circ}$ above 
the Earth limb. Typically, this was the case for segments spanning 20-38\,min
during each 97-min orbit of the spacecraft, and 
15 such segments were completed each day. The exceptions were 
three brief periods (15.67-18.13 June, 14.28-15.82 July, 12.11-13.44 August) 
when the angular distance between \tempelfull\ and the Moon was too small 
(below 15$^{\circ}$) to permit reliable pointing at the target with the 
SWAS on-board star tracker.

\begin{figure}[!p]
\begin{center}
\includegraphics[angle=0,width=4in]{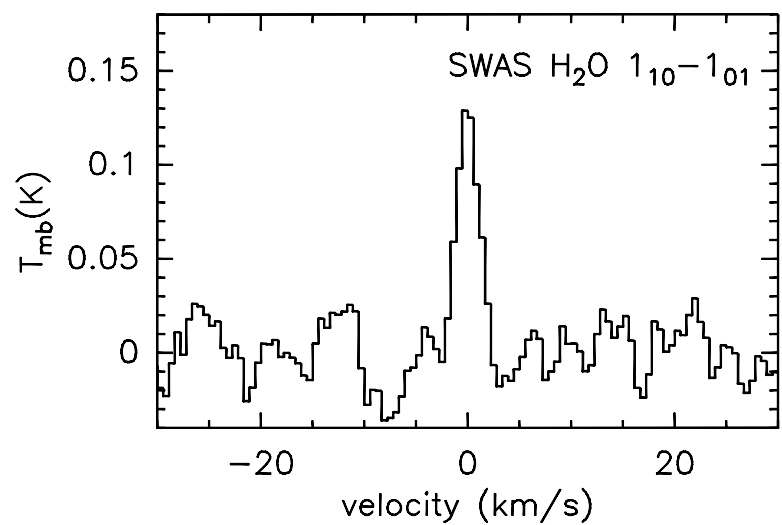}
\caption[]{SWAS-measured ground-state 556.936\,GHz 
ortho-water line toward comet \tempelfull.
The spectrum shows emission observed between 10 June and 19 June and the line
represents approximately 24\,hrs of on-source integration. The velocity scale 
of the reduced data is given in the rest-frame of the comet. 
\label{tempel1-sample}}
\end{center}
\end{figure}

The observations were conducted in position-switching mode, where observations
toward a nearby, signal-free reference position approximately 
$0.5^{\circ}$ separated from the comet were subtracted from
observations centered on the comet. This gives a total on-source 
integration time between $t_{\rm on}=$1.5-3.5\,hrs for the observations 
made over the course of one day, after subtracting the time spent 
integrating on the reference position and the time needed 
for slewing the spacecraft between on- and off-source positions
and calibration measurements. The Doppler correction for the
relative motion of the comet and the spacecraft has been made using
the ephemeris provided by JPL using their web-based 
form\footnote{JPL/SSD ephemeris available online from
http://ssd.jpl.nasa.gov/horizons.html (maintained by A.B. Chamberlin,
D.K. Yeomans, J.D. Giorgini, M.S. Keesey, P.W. Chodas, and
J. Bytof)}. The velocity scale of the reduced data is given in the 
rest-frame of the comet.  The emission 
line is partially resolved in the SWAS observations. The frequency 
resolution of the SWAS acousto-optical spectrometer (AOS) is 1.5\,MHz 
(resolution bandwidth) which corresponds to a velocity resolution of 
0.81\,\kms, and  the channel spacing of the spectrometer is 0.56\,\kms. 
Typical line widths of 
the SWAS-measured spectra range between 1.2-2.0\,\kms\ (full width at 
half maximum) with an accuracy typically between 0.25-0.45\,\kms. A line-width 
of 1.5\,\kms\ is expected for the SWAS-observations of \tempelfull\ 
assuming a spherical outgassing velocity of 0.6\,\kms\ (see section
\ref{water-prod} below). 
{\new 
The observed spectra are scaled to the main beam brightness, $\Tmb$,
using the main beam efficiency of 0.9 measured for the SWAS telescope
\citep[][]{Melnick2000}, and the velocity-integrated line intensity is 
denoted as $I_{\rm mb}=\int \Tmb\,dv$.  A sample spectrum
is shown in Fig.\ \ref{tempel1-sample}.}

\begin{figure}[!p]
\begin{center}
\includegraphics[angle=0,width=4in]{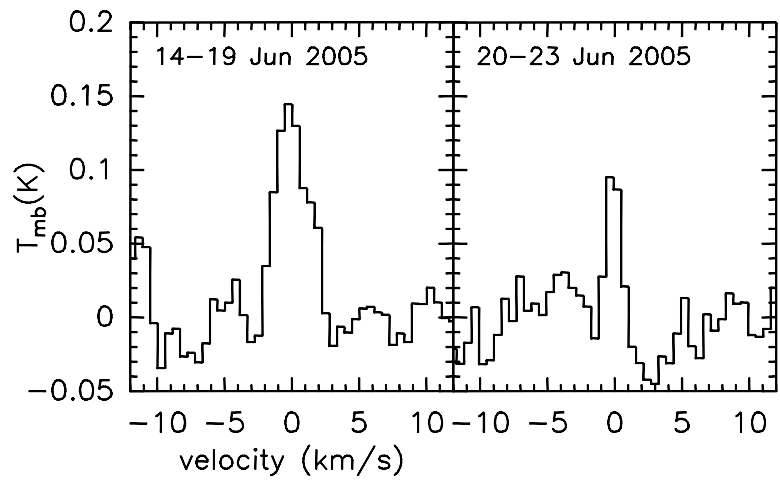}
\caption{SWAS observation of the \gs\ ground-state transition of ortho water. 
The average spectrum is shown respectively for the period of 14.03-19.99 June 
2005 (left) and 20.00-24.00 June 2005 (right). \label{min-max-spect}}
\end{center}
\end{figure}

Typically, the observations made for a period of one to three
days had to be co-added for a $5\sigma$ detection of the 
\gs\ emission of \tempelfull\ between 5 June and 10 July. The
detected line-integrated intensity varied considerably during this
period, between $\Imb=0.38$\,\Kkms and 0.11\,\Kkms, see Table 
\ref{swas-monitor-tempel1} and the spectra in Fig.\ \ref{min-max-spect}. 
The line emission was significantly weaker 
throughout the remainder of July and August. This resulted partly from the
larger comet-Earth distance and thus the larger beam dilution of the emission
for the later dates.
The comet-Sun distance, by contrast, did not change significantly over the 
course of the observations. 

The SWAS observations made on 4 July 2005 cover the (Earth received) 
time of the impact at UT 5:52:02. The corresponding orbital segment
started 29\,minutes before impact and continued until 8\,minutes
after impact. Consecutive observations of $\sim37$\,min 
duration were made every 97\,min, the approximate orbital period 
of SWAS. The on-source integration time of the observations
made during a single segment typically was $t_{\rm on}\approx$14\,min,
too short for the detection of the water vapor emission of the comet before 
impact. In addition, SWAS detected no statistical significant
emission increase from the water released by the impact, 
even when observations of several segments post-impact are co-added.

\section{Results}
\label{water-prod}
\subsection{Radiative transfer model}
We use the {\it rat4com} model by \cite{Bensch2004b} for water line 
emission in comets
to derive the total water vapor production rate
from the velocity-integrated line intensity. The 
model is based on a spherically-symmetric density profile 
for a comet with a constant outgassing rate (Haser model), and the
radiative transfer code by \cite{Hogerheijde2000}. The finite lifetime
of the water molecules due to photo-destruction in the solar UV field 
is taken into account in the model.
The photo-destruction rate for a given date was calculated from
monitoring observations of the solar continuum flux at $\lambda=10.7$\,cm 
published on the DRAO 
web-page\footnote{http://www.drao-ofr.hia-iha.nrc-cnrc.gc.ca/icarus/www/sol\_home.shtml} 
and using the relation derived by \cite{Crovisier1989}.

The water-electron collision rates have been updated for the current 
version of {\it rat4com}, using the newer calculations by \cite{Faure2004}.
The impact of the newer collision rates on the line intensity, however, 
is small for the ground-state rotational transitions ---
of the order of a few percent. Larger differences are noted only
for the transitions connecting rotationally excited levels \citep{Bensch2005}
{\new which were not observed by SWAS.}

We assume a neutral
gas kinetic temperature of $\tkin=35$\,K and an expansion
velocity of $\vexp=0.6$\,\kms\ throughout the coma. The choice
of $\vexp$ is based on the study by \cite{Bockelee-Morvan1987}
which indicates small expansion velocities of typically
$0.5\la\vexp \la0.7$\,\kms\ for weak comets ($\qwat<10^{29}$\,s$^{-1}$) 
at heliocentric distances of $\rh>1.2$\, AU. 
The kinetic temperature was calculated from
the $\tkin(\rh)$ relation derived by \cite{Biver2000}
for C/1999 H1 (Lee), a comet with a somewhat higher gas production
rate than \tempelfull. This gives a temperature of 45\, K or 25\,K,
depending on whether the power-law fit in \cite{Biver2000} to the 
comet C/1999 H1 (Lee) data for pre- or post-perihelion are used, and 
we adopt $\tkin=35\pm10$\,K for the present study. This is 
consistent with the temperatures derived from gaseous species 
observed using high-dispersion infrared spectroscopy \citep{Mumma2005}.
Our results are not very sensitive to the assumed $\tkin$ in this range,
however. The water production
rate is larger by $\la$3\% if we assume a neutral gas temperature 
of 25\,K, and smaller by less than 1\% for a higher $\tkin$ of $45$\,K.

The largest model uncertainty is the assumed electron density profile.
For the present model we use the electron density derived by the
in-situ measurements in the coma of comet 1P/Halley, 
scaled to the smaller water production rate and larger heliocentric
distance of \tempelfull\ \citep{Biver2000,Bensch2004b}.
Several studies of the \gs\ transition observed in other comets suggest 
a lower electron density (N.\ Biver, {\it priv.\ comm.}, based on 
data presented by \cite{Lecacheux2003}, and \cite{Bensch2004a}). In
the absence of mapping observations and multi-line observations 
of water rotational transitions that can be used 
to constrain the electron density, the uncertainty in the electron
density limits the accuracy of the water production rate that can be derived 
from submillimeter observations. For \tempelfull, we obtain 
a water production rate which is larger by 8-13\% if we reduce the 
electron density by a factor of two. Reducing the electron abundance by a 
factor of 10 gives a 20-25\% larger water production. The result is relatively
insensitive to the electron density assumed in the model because electrons 
dominate the rotational water-line excitation for radii between
$\sim 100$\,km and $\la 10^{4}$\,km, and thus on a linear scale which is
more than a factor of 5 smaller than the SWAS beam at the distance of the 
comet. The water-line emission detected by SWAS is 
governed by pumping by solar photons and fluorescence of water molecules in the
extended coma,
{\new accounting for 75-65\% of the emission detected by SWAS.
Much of the remaining 25-35\% can be attributed to collisions of water with
electrons, while the contribution from water-water collisions is at a level
of $\la$5\% only. Within the range indicated, the larger relative  
contribution by collisions applies to the observations made in June when 
the effect of beam-dilution of the area dominated by collisional excitation
in the SWAS beam was reduced due to the smaller comet-Earth distance.}

\subsection{Water vapor production rate June through August 2005}
\label{qwat-monitoring}
{\new The water vapor production rate of \tempelfull\ derived from
the SWAS observations is listed in Table \ref{swas-monitor-tempel1} and 
and its time evolution is shown in Fig.\ \ref{tempel1-qwater}. The error bars 
of $\sim20$\% reflect the accuracy due to the statistical 
error resulting from the baseline noise in the spectra and is larger 
than the systematic error resulting from the uncertainties in the
water collisional excitation by electrons. (While the electron density 
might be reduced in the atmosphere of the comet, it is unlikely that electron 
collisions are entirely negligible.)
For the period from June through August 2005, significant variations
are noted for the water vapor production rate 
of more than a factor of three 
on time scales as short as a few days.}
This is particularly noticeable in the results obtained in the period 
from 5 June to 10 July where the emission was strong enough to provide 
significant signal-to-noise ratios in 2 to 3 days of SWAS observations.
For example, the maximum of $\qwat=(12.9\pm2.5)\times 10^{27}$\,s$^{-1}$
(average for 14.0-15.7 June) is followed by a minimum in the 
water production rate of
$\qwat=(3.8\pm1.4)\times 10^{27}$\,s$^{-1}$ only $\sim 5$ days later.
{\new Generally, the water vapor production rate fluctuates with
periods of increased production rates ($\qwat \sim 10^{28}$\,s$^{-1}$) 
alternating with periods of reduced activity 
($\qwat \la 5\times 10^{27}$\,s$^{-1}$).
In particular,  three episodes of increased activity are noted
between June to mid-July 2005.}

We checked for periodic signals in the SWAS data resulting from the 
rotation of the nucleus by co-adding the data by phase. However, we found 
that in order to resolve any periodicity in phase-space, the 
signal-to-noise in the co-added bins is insufficient to obtain meaningful
results. Thus, it is not possible to decide whether or not the 
SWAS-observed line intensity was also periodic on timescales reported
by other observers.

The comet activity continued to be highly variable after 10 July.
The weaker line emission owing to the larger distance to the comet
reduced the signal-to-noise ratio in the spectra and thus the time-resolution 
of the SWAS-measured water production rate as more spectra had
to be co-added to detect the signal. While an emission line is 
detected in the co-adds of 2-3 days during phases of increased activity,
the emission from the comet between these peaks was either at or below
the SWAS detection limit even for co-adds of up to $\sim$10 days.

The water vaporization rate of the nucleus is 
a factor of two lower post-perihelion than pre-perihelion, if the 
average over longer periods ($\ga 25$ days) is considered (last
three entries in Table \ref{swas-monitor-tempel1}).
Such asymmetries are also evident in the observed light curves for 
\tempelfull\ made during previous apparitions of
the comet \citep[for example,][]{Meech2005b}, 
as well as for the perihelion passages of other comets 
\citep{Biver2000,Chiu2001}, and in particular Jupiter-family comets 
\citep{AHearn1995}. 
{\new Based on a composite light curve from observations made 
before 2005, \cite{Lisse2005} find that \tempelfull\
reaches its maximum activity approximately 60 days pre-perihelion.}

The SWAS monitoring observations give a total of 
$N\approx 4.5\times 10^{34}$ water molecules ($\approx 1.3\times 10^9$\,kg) 
released by the comet from June through August 2005. {\new Considering that
the peak activity might have occurred 60\,days before perihelion, and thus
$\sim30$ days before the start of the SWAS monitoring observations,
the total amount of water lost by \tempelfull\ during the 2005 apparition
might have well been $\sim 10^{35}$ water molecules 
($\sim 2.9\times 10^9$\,kg, approximately 
twice the amount measured by SWAS for the period from 5 June to 
1 September 2005. We estimate a net erosion of $\sim 43$\,cm for the 
active surface areas for the 2005 perihelion passage, ignoring the 
contribution from dust and gaseous species other than water and assuming 
the nuclear properties summarized by \cite{AHearn2005} and \cite{Belton2005}
with $\sim$9\% of the $113$\,km$^2$ total surface area being active,
and a mean density of 0.6\,g\,cm$^{-1}$. 
Alternatively, if the
scarps identified on the nucleus are the areas responsible for
much of the comet's activity, the amount of material released
causes the scarp to retreat by $\sim$220\,m per perihelion
passage. (Assuming that the scarps have a height of 20\,m and a
total length of one kilometer).}

\subsection{Observations following the impact}
\begin{figure}[!p]
\begin{center}
\includegraphics[angle=0,width=4in]{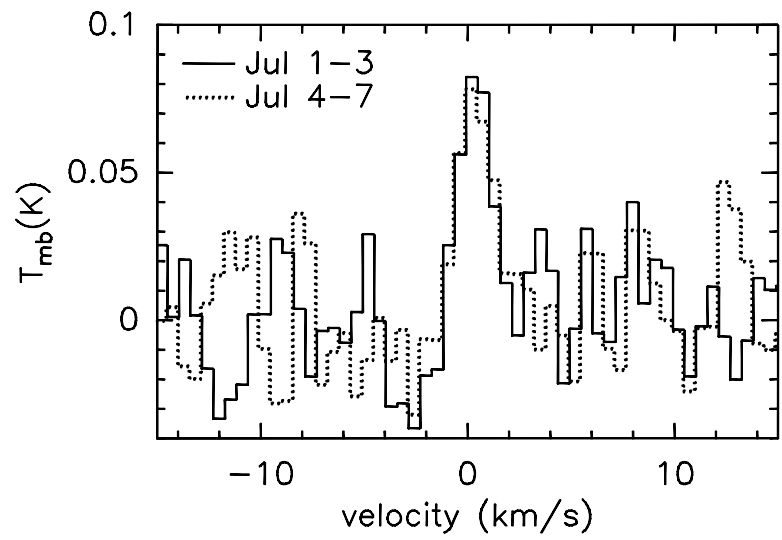}
\caption{SWAS spectra of \tempelfull, respectively, averaged over a period 
of 3 days before and after the impact. \label{before-after}}
\end{center}
\end{figure}

No increase in the water vapor emission resulting from the impact on 
4 July was detected by SWAS, either as a short-lived outburst in intensity 
due to vaporized water ice in the ejected material
or as a permanently increased water vapor production rate from a newly formed
active area. In fact, the velocity-integrated line intensity
averaged over the three days before and after impact is 
virtually identical ($\Imb=0.16\pm0.04$\,K\,\kms\ for both dates; see Fig. 
\ref{before-after} and Table \ref{swas-monitor-tempel1}). 
This corresponds to a water vapor production rate
of $\qwat=(6.6\pm1.5)\times10^{27}$\,\pers\ for 1-3 July, and 
$\qwat=(6.9\pm1.6)\times10^{27}$\,\pers\ for 4-6 July.
The larger production rate for the later date results from the somewhat
larger comet-Earth distance (beam-dilution of the emission) that requires 
the comet to have a slightly larger $\qwat$ to give rise to a spectral line 
with the same integrated intensity.
The water vapor production rate derived from a 3-day co-add
of SWAS observations centered on the impact date is 
$(5.2\pm 1.5)\times10^{27}$\,\pers, consistent with the 3-day co-adds
before and after the impact. This SWAS-measured water production rate
is a factor of two lower than that derived by \cite{Mumma2005} from 
their pre-impact observations of water hot-band
emission in the near infrared, but is consistent with the 
water production rate measured by OH observations at UV wavelengths 
\citep{Kueppers2005,Schleicher2006}. Published water production rates 
for \tempelfull\ from other observations of submillimeter
water lines were not yet available at the time of writing.

In order to derive an upper limit on the water vapor ejected by the impact,
we extended the radiative transfer model to include comets for which the water 
production is time-dependent. The total water vapor production rate is then 
replaced by $\qwat=Q_{\rm q}+Q_{\rm b}(t)$, where 
$Q_{\rm q}$ is the constant component. For \tempelfull\ before impact we adopt
$Q_{\rm q}=5.2\times 10^{27}$\,\pers.
$Q_{\rm b}(t)$ is the time-variable component representing the water
vapor produced by the impact, assumed to be a box-car function
for the present simulations. In this case the water production 
rate is elevated (but constant) for the duration $\tau$ of the outburst
and returns to the pre-impact level $Q_{\rm q}$ afterwards. The total
number of water molecules released by the impact in our model is 
$N_{\rm model}=Q_{\rm b}\times \tau$. A more detailed description of 
the model for water vapor rotational-line emission in comets undergoing 
outbursts will be presented in a future paper (Bensch {\it et al., in prep.}).

\begin{figure}[!p]
\begin{center}
\includegraphics[angle=-90,width=5.5in]{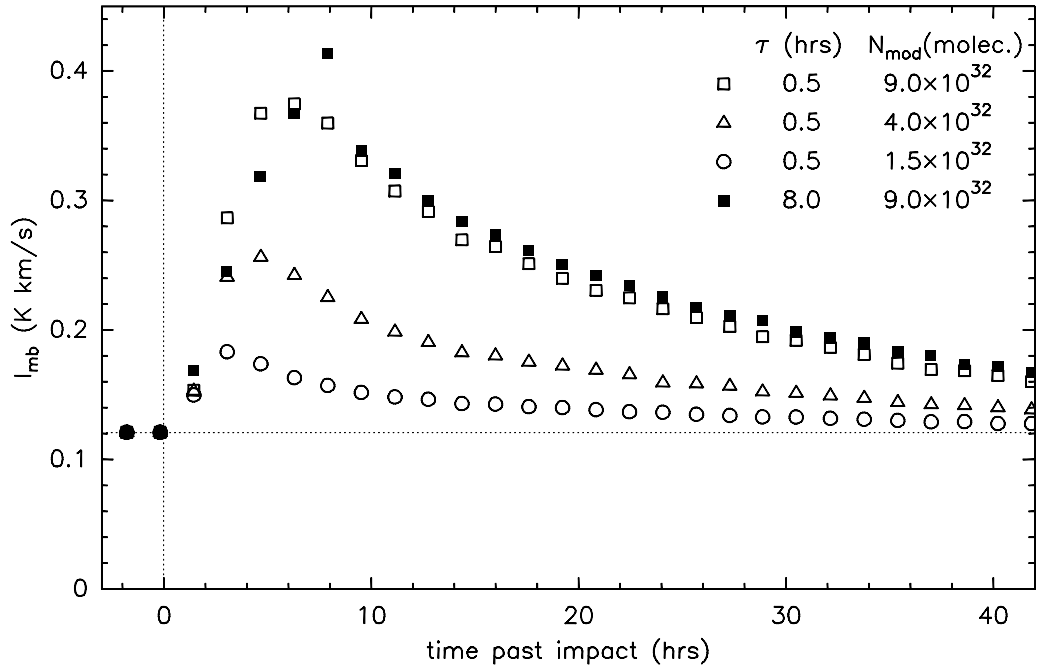}
\caption{Velocity-integrated intensity calculated for four different
outburst models (see legend at top right). Open symbols show three models 
with an outburst duration
of $\tau=$30\,min but a different amount of water released by the outburst. 
The filled symbols give the results for a model with $\tau=8$\,hrs. 
The pre-outburst level $I_{q} = 0.121$\,K\,km\,s$^{-1}$ is indicated by the 
dotted horizontal line. \label{outburst-simulations} }
\end{center}
\end{figure}

Simulations for SWAS observations of an outburst following the
impact are presented in Fig.\ \ref{outburst-simulations}. 
The figure shows the time evolution of the line-integrated intensity for 
four different outburst models, three models with $\tau=30$\,min 
($N_{\rm model}= 1.5, 4,$ and $9\times 10^{32}$ water molecules), and one 
model with $\tau= 8$\,hrs  ($N_{\rm model}=9\times 10^{32}$ water molecules). 
The time step in the model series reflects  the timing of the SWAS observations
(orbital segments) after the impact. The model results presented in the 
figure show that the signal 
in the first few hours after the impact is insensitive to the total number of 
water molecules released. During this phase the ejected water vapor 
fills only a small portion of the SWAS radio beam, and optical 
depth effects play an important role. At later times $t\gg\tau$, when the 
spherical outflow has distributed the gas over a larger area, the
beam-averaged optical depth has decreased and the excess emission resulting
from the outburst scales approximately linearly (within $\la 10$\%) 
with the number of molecules for the models. This linear behavior
is evident from a comparison of the models shown by the open symbols in
Fig.\ \ref{outburst-simulations} for times $t\ge 6.3$\,hrs.
By contrast, the line intensity predicted by the model for $t\gg\tau$ is 
relatively insensitive to the duration of the outburst.

\begin{figure}[!p]
\begin{center}
\includegraphics[angle=0,width=4in]{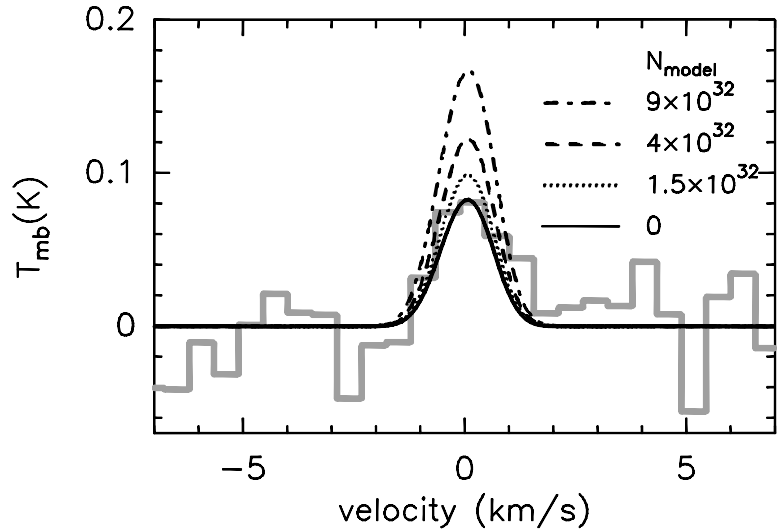}
\caption{Comparison of SWAS observations and model results for the post-impact
period of 4.3 to 6.0 July. The SWAS-observed spectrum 
is shown as a histogram (weighted average, see text for details).
The dashed and dotted lines give the results for 
outburst models with $\tau=30$\,minutes but three different 
amounts of water in the ejected material.
For comparison, the signal expected for a model with
a constant water production rate of $\qwat=5.2\times10^{27}$\,\pers\
is shown as a solid line. The latter is the water production rate of the
comet averaged for a 3-day period centered on the impact date.
\label{outburst-upper-limit}}
\end{center}
\end{figure}

For a comparison of the SWAS
observations to the model, we fit a Gaussian line profile to the 
observed spectra in each orbital segment after the impact, where we have
constrained the fit using the position and width of the line 
in the model spectrum and only allowed the line intensity to vary.
This procedure gives us a line-integrated intensity and a statistical error
for the observations made in each orbital segment $i$, 
$I_{\rm obs,i}\pm \Delta I_{\rm obs,i}$.
Note, that no emission line is detected in the spectrum of an 
individual segment and that the error $\Delta I_{\rm obs,i}$ generally is 
larger than the integrated intensity from the fit.
However, a spectral line of integrated intensity consistent with the 
pre-impact intensity is detected if 
$\ga 25$ segments are co-added.  (Fig.\ \ref{outburst-upper-limit}).

For the evaluation of the {\new post-impact} SWAS data,
we included observations made up to 42\,hrs after the impact.
In order to derive an upper limit to the water released by the impact
we calculate a weighted average of the
SWAS observations, where the spectra in each segment are weighted
by the excess intensity resulting from the outburst predicted by the
model ($I_{\rm model}-I_{\rm q}$),
and by the inverse square of the 
statistical error of the observed line intensity $(\Delta I_{\rm obs})^{-2}$.
The weighted average of the SWAS observations made in several segments 
allows us to derive a better upper limit than for a single segment, and at the
same time avoids that segments where the model predicts a lower
intensity are co-added with an equal weight.
For a model with a given $\tau$ and $N_{\rm model}$ we first 
estimate the number of water molecules in the outburst for each individual
segment $i$ post-impact by $N_i = 
N_{\rm model}\,(I_{\rm obs,i}-I_{\rm q})/(I_{\rm model,i}-I_{\rm q})$.
The accuracy $\Delta N_i$ is estimated by standard error 
propagation of the statistical error of the observed intensity, 
$\Delta I_{\rm obs,i}$, and of the line intensity before impact, 
$\Delta I_{\rm q}$. The final estimate for the number of water molecules 
in the outburst of duration $\tau$ is then calculated 
as the weighted average of the $N_i$,
\begin{displaymath}
N = \frac{\sum_i N_i/(\Delta N_i)^2}{\sum_i (\Delta N_i)^{-2}},
\end{displaymath}
and the accuracy of this estimate is given by 
$\Delta N = (\sum_i (\Delta N_i)^{-2})^{1/2}$. The summation is over 
the orbital segments $i$ after impact. 
{\new Note, that with 
$\Delta I_{\rm obs,i}> \Delta I_{\rm q}$ we have 
$\Delta N_i\approx \Delta I_{\rm i,obs}/N_{\rm model}\,
(I_{\rm model,i}-I_{\rm q})$ and thus the individual $N_i$ are weighted
by $(\Delta N_i)^{-2} \approx 
(I_{\rm model,i}-I_{\rm q})^2/(\Delta I_{\rm i,obs})^{2}$.
}

{\new The outlined procedure} assumes that the excess emission from the
outburst is proportional to the number of water molecules in the outburst,
whereas optical depth effects mentioned above generally result in a  
non-linear relation between the number of water molecules in the coma
and the detected line intensity. In practice, the upper limit derived 
from the observations is checked by comparing the results for models 
with different $N_{\rm model}$, or an iterative process, if needed.
%
%

{\new
The duration of the outburst following the impact is not
known. However, observations suggest that it was relatively short,
probably less than a day \citep{Kueppers2005}.
Assuming an outburst duration $\tau=30$\,minutes we 
obtain $N = (-1.9\pm 2.6)\times10^{32}$ water molecules for the SWAS
observations post impact, and thus a $3\sigma$
upper limit of $5.9\times 10^{32}$ water molecules. 
Fig.\ \ref{outburst-upper-limit} compares a weighted average of the SWAS 
data post-impact and the model results for four different $N_{\rm model}$. 
Table \ref{water-upper-limits} lists the results (3$\sigma$ upper limits) for 
three additional models where we assumed a longer duration for the
outburst $\tau=2,8,16$\,hrs.}
The line intensity remains elevated for a longer period in outbursts
with $\tau>2$\,hrs and the observations of more segments are co-added
in this case, resulting in a slightly lower upper limit for the water vapor 
released by the impact. Therefore, $N<6\times 10^{32}$ {\new molecules}
($\sim 1.8\times 10^7$\,kg) is a conservative upper limit to the 
water vapor released by the impact.

\section{Discussion}
\label{discuss}
No increase in the water rotational-line emission from the impact ejecta 
was detected by SWAS and we derive a $3$ sigma upper limit to the water 
vapor generated by the impact, $N_{\rm tot}<6\times10^{32}$ water molecules
($1.8\times10^7$\,kg). The SWAS non-detection is consistent with  
results from OH observations at UV wavelengths by \cite{Kueppers2005} 
which indicate that only a small amount of water vapor 
($\sim 4.5\times 10^6$\,kg) was produced by the impact and suggest 
that the dust-to-water ice mass ratio of the ejected material
exceeded unity \citep{Keller2005}. However, care has to be taken in 
arriving at this conclusion.  Ice freshly excavated from the impact 
crater might sublimate rather slowly on exposure to sunlight
{\new if the ice was ejected in large chunks}, whereas dust 
ejected into escape orbits by the impact would only linger near the comet 
for short periods.  Early post-impact observations could thus provide 
a misleading impression of the longer-term integrated dust to water ratio 
lost by the comet in response to the impact.

No newly formed active area was detected by SWAS to 
significantly contribute to the water vapor production rate of the comet 
for more than 2-3\,days. The non-detection of water vapor due to the 
Deep Impact event argues against the impact of meteoritic material as 
the origin of the natural outbursts observed in \tempelfull. This is 
suggested by the small amount of water released by the impact when 
compared to the natural outbursts, and by the fact that this would
require an abundance of meteorites in interplanetary space far
higher than all other observations permit in order to explain the
number of outbursts observed in \tempelfull\ {\new
\citep[eg.,][and references therein]{Lang1992,Ceplecha1992,Greenberg1994}. }
{\new Another argument against exogenic processes, such as meteorite
impacts, as the reason for the outbursts is the correlation of the 
outbursts with the light curve observed by \cite{AHearn2005}.} 

\begin{figure}[!p]
\begin{center}
\includegraphics[angle=-90,width=5.5in]{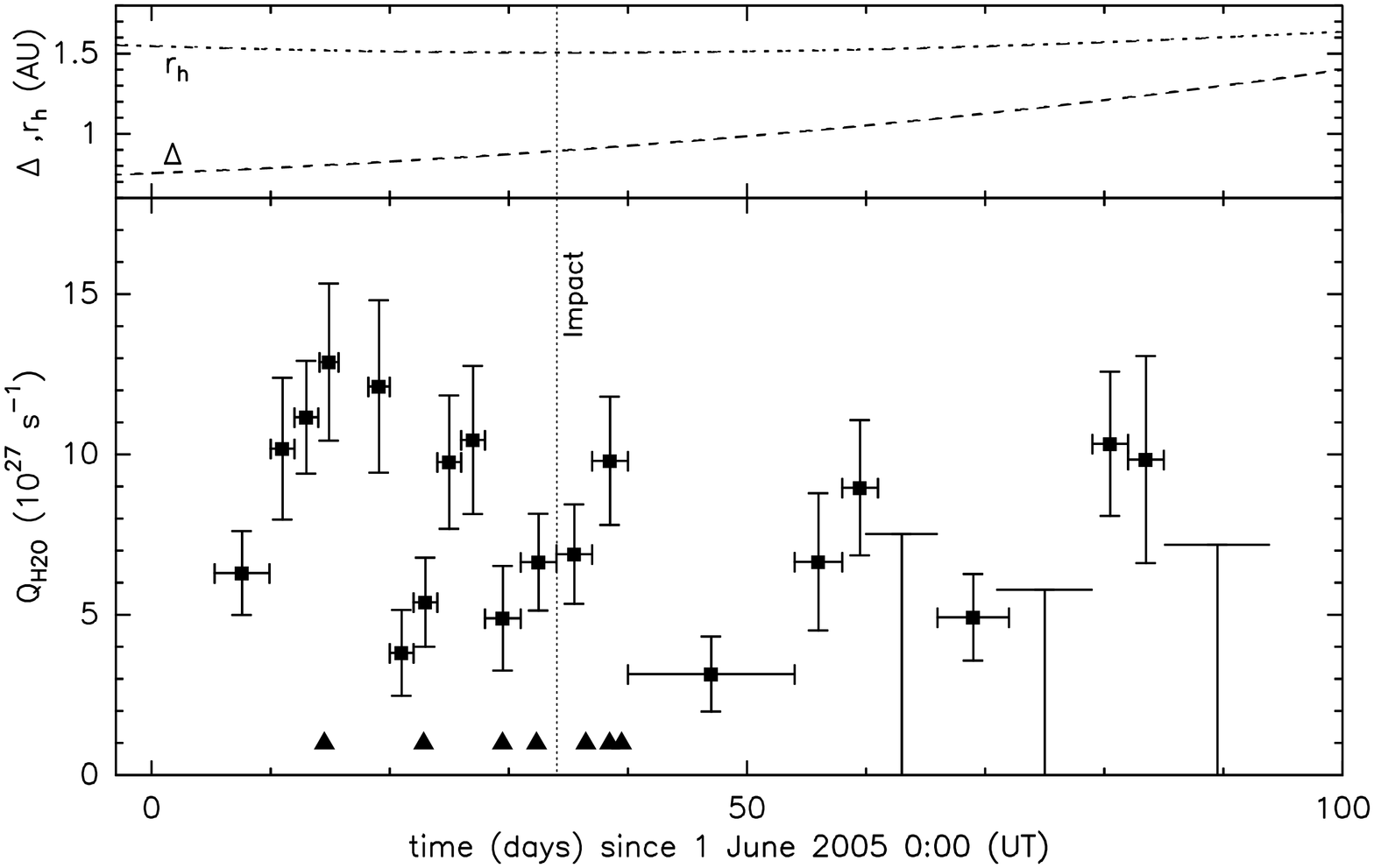}
\caption{Top panel: Comet-Sun distance ($\rh$) and Comet-Earth distance 
($\Delta$) for \tempelfull\ for the period of June through August 2005. 
Bottom panel: total water production rate $\qwat$ derived from SWAS 
observations of the \gs\ transition of ortho water. The impact date is
marked by the vertical line. Outbursts reported from observations at 
other wavelengths are marked by filled triangles \citep{Meech2005a}. 
Note, that the clustering of open triangles
in June and July 2005 does not necessarily indicate 
that the comet was more active during this period. Rather, it likely 
results from observational bias, since most of the observing campaigns for 
\tempelfull\ had been conducted during the days leading up to
and immediately following the Deep Impact encounter with the comet on 4 July.
\label{tempel1-qwater}}
\end{center}
\end{figure}

\begin{figure}[!p]
\begin{center}
\includegraphics[angle=-90,width=4.5in]{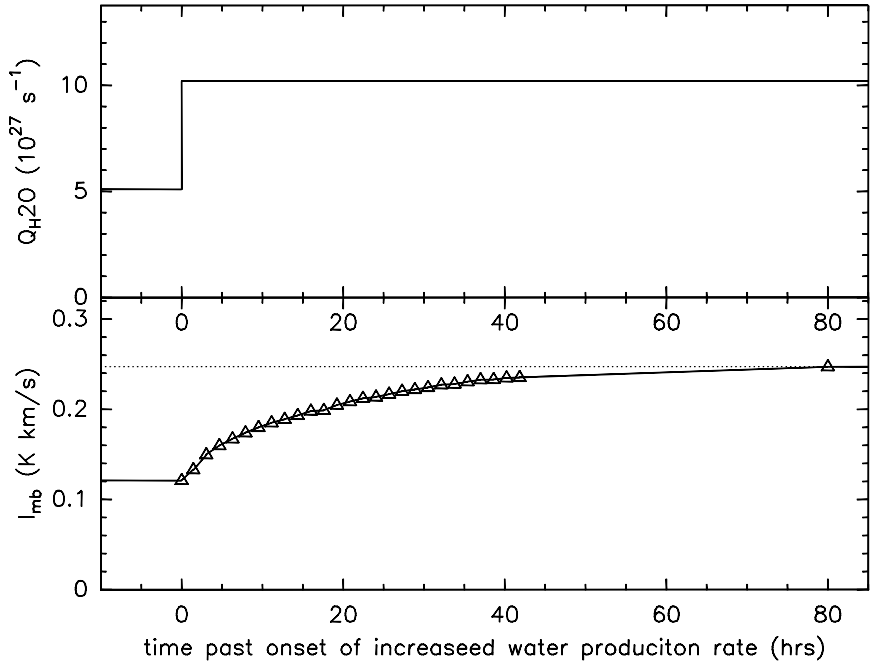}
\caption{Response of the SWAS-detected line intensity (bottom panel)
to an instant, two-fold increase of the water production rate (top panel). 
The lower panel shows the evolution of the line-integrated intensity if the 
water production rate of the comet is increased from 5.2 to 
10.4$\times10^{27}$\,s$^{-1}$ at $t=0$. (The vertical lines give the 
line intensity for a steady-state model with these water production rates).  
This model is applicable to a situation where a new active area is created,
for example, following an impact or a splitting of a comet nucleus.
Other model parameter ($\rh, \Delta, \vexp$) are those applicable 
to \tempelfull\ on 4 June 2005. 
\label{double-water-prod}\label{lastfig}}
\end{center}
\end{figure}

{\new
The water production rate of \tempelfull\ shows large natural variations, 
and the observed fluctuations are somewhat surprising and difficult 
to explain.  While the surface material of the comet is largely desiccated, 
periodic spurts of water vapor emission, nevertheless, are observed. 
Strictly, a model similar to the outburst model presented for the observations 
made during the Deep Impact event has to be employed for the water monitoring
observations, particularly if one considers that the natural outbursts 
and the overall variability of the water outgassing rate of the
comet is much larger than the amount of water vaporized by the impact.
While the present paper for the first time presents a model
for radio-line emission in outbursting comets, such a modeling is currently 
computationally prohibitive for the entire data set 
of the \tempelfull\ observations unless the timing of the
outbursts are known (the outburst duration and start time). 
Several of the strong outbursts recorded by Earth-based instruments
are summarized by \cite{Meech2005a}, but 
it is not always possible to exactly constrain the timing of these outbursts 
because of the limited time coverage of these observations and because tracers 
other than water were observed in most cases. 
In fact, the opportunity to correlate observed outburst signals with a 
well-defined impact date was one of the reasons to conduct the Deep Impact 
experiment.
 
For the SWAS observations of a natural outburst, because we average over 2-3 
days, it is not possible to decide whether the water molecules were released 
in a relatively short time ($\tau\ll 3$\,days) or at a lower rate over a 
longer period ($\tau\la$3\,days).
Presenting an average $\qwat$ assuming a constant water production rate 
during the period where the SWAS-detected signal was averaged
is therefore acceptable to trace the variation of the cometary activity 
on time-scales of $\ga$days. 
Since SWAS is sensitive to the total number of water molecules in the 
extended coma where line excitation is governed by IR-fluorescence,
a sudden increase in the water production rate is not immediately visible
in the SWAS signal but increases gradually over 1-2 days 
(Fig.\ \ref{double-water-prod}).
Assuming a two-fold increase of the water production rate, the SWAS-detected 
signal has reached 90\% of the line intensity of the corresponding
steady-state model with twice the water production rate within 
$\sim30$\,hours.
Thus, one has to note that the SWAS-derived water production
rate for \tempelfull\ presented with Table \ref{swas-monitor-tempel1} and Fig.\
\ref{tempel1-qwater}) might lag the true water production rate by 
$\sim 1.3$\,days during phases of strong comet variability.
A similar conclusion applies to all studies of the water 
production rate using water-line observations covering a significant fraction 
of the gaseous coma using a Haser model. 
}

The main indication of desiccation comes from the high surface temperatures 
and the associated low thermal conductivities and inertia on the sunlit 
side of the comet.  \cite{AHearn2005} find that \tempelfull\ has a largely 
homogeneous color and  uniform albedo as low as $0.04 \pm 0.02$ over its 
entire surface. At perihelion, maps of the nucleus show temperature
variations across the surface between $260 \pm 6$ K to $329 \pm 8$ K, 
with an absolute calibration uncertainty of order 20\%. Even at the lowest 
cited temperature, however, the sublimation rate of ice close to the comet 
surface would be 
{\new significantly}
higher than observed.  
As \cite{AHearn2005} also point out, the high surface temperature argues 
for a thermal inertia probably lower than 
$10^5$ erg K$^{-1}$ cm$^{-2}$ s$^{-1/2}$.  This implies that the 40.83\,hr
periodic heat pulse on the sunlit side corresponding to the observed 40.83\,hr 
rotation period observed close to perihelion does not penetrate deep into the 
comet's surface, and is consistent with the low intra-pulse evaporation rate 
of water.  {\new Most of} the frozen water apparently resides deeper 
beneath the surface than the heat pulse penetrates, and remains at a 
temperature sufficiently low to limit sublimation. 
{\new \cite{Sunshine2006} detect water ice on the surface of
\tempelfull, but these ice deposits cover too small a fraction of the 
surface to account for the total water vaporization rate observed 
for the comet.}
 
A less compelling indication for dehydration of the comet's surface layers 
comes from the debris released by the July 4 impact. \cite{AHearn2005} 
conclude that ejecta consisted largely of very fine dust in the 1 to 100 
micrometer size range, while \cite{Meech2005a} find the ratio of dust to gas 
in the impact ejecta to have been significantly higher than in the pre-impact 
emission of material from the comet.  
\cite{Meech2005a} 
show that the light curve mainly associated with dust emission rose to a 
maximum within roughly half an hour after impact and then declined after 
two hours, \cite{Mumma2005} estimate a doubling of the post-impact water 
vapor production rate integrated over a 24 hour period. 
A contribution to the water vapor production over this 
longer period could have come from newly excavated material ejected from the 
crater and strewn over a sizeable surrounding area, consistent with the 
observation that an appreciable fraction of the excavated material fell back 
onto the comet surface.  The ice in the ejecta would subsequently have been 
sublimated by sunlight.  As \cite{Kueppers2005} point out, the heat required 
for the evaporation of this much water could not have been provided by the 
energy of impact and would instead have had to come from the Sun.

\cite{AHearn2005} point out that a number of pre-impact cometary
outbursts appear to be correlated with the comet's rotation rate. 
More specifically, 
they identify {\new emission from a topographic feature, possibly a scarp
or depression,} that sees the sun rising around the time of each outburst.  
The sudden heating of the surface of the {\new topographic feature} by direct 
sunlight could cause both rapid sublimation and the release of dust 
grains no longer held together by ice. Smaller grains could be carried 
off into space by the evaporating water.
The gravitational attraction of the comet is sufficient to have retained a 
significant fraction of  the solid impact ejecta, as shown by the plume of
material raining back down onto the comet's surface after impact. 

The SWAS water vapor detection history shown in Fig.\ \ref{tempel1-qwater} 
indicates a rough correlation between the times of a natural outburst 
identified by other observers, and a rise in the 557\,GHz water vapor emission 
integrated over a day or two following the outburst. {\new This time delay
is consistent with the expected response of the SWAS-detected signal to an
instant increase of the water vaporization rate of the nucleus shown
in Fig.\ \ref{double-water-prod}.}
{\new In this context it should be noted that the outbursts indicated in
Fig. \ref{tempel1-qwater} are ``big'' natural outburst which were
also detected by Earth-based telescopes, while the weaker outbursts
reported by \cite{AHearn2005} were too faint to be detected by Earth-based
instruments. }

If the prime source of water vapor is 
{\new a tilted topographic feature on the surface such as a scarp or wall}, 
and the outburst occurs at sunrise, a slight slowing down of the comet's 
rotation should become 
evident in the course of time.  The momentum of evaporating water along the 
direction of rotation is $\sim N(mkT)^{1/2}$, where $N$ is the total number 
of water molecules released over a period of time, $T\sim 300$ K is the surface
temperature at sunrise and $m$ is the molecular weight of water. With a
lower limit to the number of water molecules vaporized from the nucleus 
during the 2005 perihelion passage of  {\new $N\sim 10^{35}$ } 
(Sect.\ \ref{qwat-monitoring}), the net angular momentum transferred 
to the comet nucleus through evaporation at radius $r \sim 3$ km should 
have been roughly $N(mkT)^{1/2}r\sim 3\times 10^{22}$g cm${^2}$ s$^{-1}$.
The angular momentum  of the comet is $\sim \pi Mr^2/P$, 
where $P\sim 40.83\pm0.33$ hr is the rotation {\new period}, 
$M\sim 4\pi \rho r^3/3$ and $\rho\sim 0.6$\,g\,cm$^{-3}$ is the 
density of the nucleus cited by \cite{AHearn2005}. For comet \tempelfull, 
$\pi Mr^2/P\sim 2\times 10^{23}$\,g\,cm$^2$ s$^{-1}$. 
While the fractional mass lost by the comet due to the sublimation of ice
is quite small, the velocity of sublimated molecules is several thousand
times higher than the rotational surface velocity of the comet. A systematic
change in the rotational period of the comet could, therefore, amount to as
much as {\new several} percent during a single perihelion passage.

Whether a slowdown in 
the comet's rotational period by up to a few percent during the 2005 
perihelion passage could be culled out from the data is not clear.
\cite{Belton2005} give the pre-perihelion rotation period of the comet at
heliocentric distances greater than 4 AU as $41.85\pm 0.10$ hours, 
although it is not
clear whether this significantly differs from the 40.83 hours measured at
perihelion. If so, the comet's rotation rate appears to have sped up during
perihelion passage. Either way, it would be useful to re-measure the
post-perihelion rotation period at large heliocentric distances to see
whether significant changes in the rotation period of Tempel 1 can be
established.

\ack
Support of the SWAS mission is provided by NASA through SWAS contract 
NAS5-30702. Additional support for the SWAS 9P/Tempel 1
observing campaign was obtained by the Deep Impact team. The help and 
cooperation of NASA HQ is greatly appreciated.

\label{lastpage}





\clearpage	

{
\renewcommand{\baselinestretch}{1}
\small\normalsize

\begin{table}
\begin{center}
\textbf{Observations and Water Vaporization Rate}
\begin{tabular}{lllll}
\hline 
\hline
Date      &  $r_h$ & $\Delta$ & $I_{\rm mb}=\int T_{\rm mb}\,dv$ & $\qwat$ \\
          &  (AU)  & (AU)     & (K\,km\,s$^{-1}$)     & ($10^{28}\,s^{-1}$) \\
\hline
Jun\,5-9   & 1.53   & 0.78     & $0.193\pm0.041$  &  $0.63\pm0.13$ \\
Jun\,10-11 & 1.53   & 0.79     & $0.304\pm0.067$  &  $1.02\pm0.22$ \\
Jun\,12-13 & 1.52   & 0.80     & $0.334\pm0.053$  &  $1.12\pm0.18$ \\
Jun\,14-15 & 1.52   & 0.81     & $0.382\pm0.073$  &  $1.29\pm0.25$ \\
Jun\,18-19 & 1.52   & 0.82     & $0.352\pm0.079$  &  $1.21\pm0.27$ \\
Jun\,20-21 & 1.51   & 0.83     & $0.106\pm0.039$  &  $0.38\pm0.13$ \\
Jun\,22-23 & 1.51   & 0.84     & $0.149\pm0.041$  &  $0.54\pm0.14$ \\
Jun\,24-25 & 1.51   & 0.85     & $0.273\pm0.059$  &  $0.98\pm0.21$ \\
Jun\,26-27 & 1.51   & 0.86     & $0.286\pm0.064$  &  $1.05\pm0.23$ \\
Jun\,28-30 & 1.51   & 0.87     & $0.127\pm0.044$  &  $0.49\pm0.16$ \\
Jul\,1-3   & 1.51   & 0.89     & $0.163\pm0.038$  &  $0.66\pm0.15$ \\
Jul\,2.7-5.7 & 1.51 & 0.89     & $0.121\pm0.036$  &  $0.52\pm0.15$ \\
Jul\,4-6   & 1.51   & 0.90     & $0.164\pm0.038$  &  $0.69\pm0.16$ \\
Jul\,7-9   & 1.51   & 0.92     & $0.221\pm0.050$  &  $0.98\pm0.20$ \\
Jul\,10-23 & 1.51-1.52 & 0.93-1.01 & $0.067\pm0.022$  &  $0.32\pm0.12$ \\
Jul\,24-27 & 1.52      & 1.01-1.03 & $0.133\pm0.044$  &  $0.67\pm0.21$ \\
Jul\,28-30 & 1.53      & 1.03-1.07 & $0.169\pm0.041$  &  $0.90\pm0.21$ \\
Jul\,30-Aug\,4 & 1.53-1.54 & 1.05-1.10 & $<0.135$
                                        &  $<0.75$\\
Aug\,5-10  & 1.54      & 1.10-1.14 & $0.084\pm0.024$  &  $ 0.49\pm0.14$ \\
Aug\,10-17 & 1.55-1.57 & 1.14-1.20 & $<0.092$
                                        &  $<0.58$\\
Aug\,18-20 & 1.57-1.58 & 1.20-1.23 & $0.154\pm0.034$  &  $1.03\pm0.23$ \\
Aug\,21-23 & 1.58      & 1.23-1.25 & $0.125\pm0.042$  &  $0.98\pm0.32$ \\
Aug\,24-Sep\,1 & 1.59-1.61 & 1.25-1.35 &  $<0.094$
                                            &  $<0.72$\\
\hline
Jun\,5-Jul\,3  & 1.53-1.51 & 0.78-0.89 & $0.225\pm0.017$ & $ 0.81\pm0.06$ \\
Jul\,10-Aug\,7 & 1.51-1.54 & 0.93-1.11 & $0.065\pm0.017$ & $ 0.34\pm0.08$ \\
Aug\,8-Sep\,1  & 1.54-1.61 & 1.12-1.35 & $0.074\pm0.015$ & $ 0.52\pm0.10$ \\
\hline
\end{tabular}
\caption[]{\label{swas-monitor-tempel1}
SWAS monitoring and water vaporization rate of comet \tempelfull. In
columns {\new 4 and 5, the error bars are one sigma} and upper limits are 
3 sigma.}
\end{center}
\end{table}

\clearpage

\begin{table}
\begin{center}
\textbf{Total Water released by the Impact}\\
\begin{tabular}{cc}
\hline 
\hline
Burst Duration & Upper limit \\
(Model)        &           \\
 $\tau$ (hrs)     & $N$ (molec.) \\
\hline       
0.5      & $<  5.9\times 10^{32}$ \\
2.0      & $<  6.1\times 10^{32}$ \\
8.0      & $<  5.3\times 10^{32}$ \\
16.0     & $<  3.9\times 10^{32}$ \\
\hline
\end{tabular}
\caption[]{
\label{water-upper-limits}\label{lasttable}
SWAS-derived upper limits on the water released by the impact.
The $3 \sigma$ upper limit is given for the assumed duration $\tau$ of the
outburst}
\end{center}
\end{table}

}

\clearpage


\clearpage

\clearpage

\clearpage

\clearpage

\clearpage

\clearpage

\end{document}